\documentclass[useAMS,usenatbib]{mn2e}

\usepackage{verbatim,graphicx,epsfig,dcolumn}
\newcolumntype{.}{D{.}{.}{4}}
\newcolumntype{,}{D{.}{.}{2}}
\newcolumntype{;}{D{.}{.}{1}}
\newcommand{\nodata}{$\cdot\cdot\cdot$}
\newcommand{\lesssim}{{\lower-1.2pt\vbox{\hbox{\rlap{$<$}\lower5pt\vbox{\hbox{$\sim$}}}}}}
\newcommand{\gtrsim}{{\lower-1.2pt\vbox{\hbox{\rlap{$>$}\lower5pt\vbox{\hbox{$\sim$}}}}}}

\title[EU Del]{EU Del: exploring the onset of pulsation-driven winds in giant stars}
\author[I. McDonald et al.]{I.~McDonald$^{1}$\thanks{E-mail: mcdonald@jb.man.ac.uk}, A.~A.~Zijlstra$^{1}$, G.~C.~Sloan$^{2}$, E.~Lagadec$^{3}$, C.~I.~Johnson$^{4}$, \newauthor
S.~Uttenthaler$^{5}$, O.~C.~Jones$^{1,6}$, C.~L.~Smith$^{1,7}$\\
$^{1}$Jodrell Bank Centre for Astrophysics, Alan Turing Building, Manchester, M13 9PL, UK\\
$^{2}$Cornell Center for Astrophysics and Planetary Science, Cornell University, Ithaca, NY 14853-6801, USA\\
$^{3}$Observatoire de la C\^ote d'Azur, Boulevard de l'Observatoire, CS 34229, F 06304 Nice Cedex 4, France\\
$^{4}$Harvard--Smithsonian Center for Astrophysics, 60 Garden Street, MS--15, Cambridge, MA 02138, USA\\
$^{5}$University of Vienna, Department of Astrophysics, T\"urkenschanzstra{\ss}e 17, 1180 Vienna, Austria\\
$^{6}$STScI, 3700 San Martin Drive, Baltimore, MD 21218, USA\\
$^{7}$Center for Research in Earth and Space Science (CRESS), York University, 4700 Keele Str, Toronto, ON M3J 1P3, Canada\\
}

\begin{document}

\date{Accepted 9999 December 32. Received 9999 December 32; in original form 9999 December 32}

\pagerange{\pageref{firstpage}--\pageref{lastpage}} \pubyear{9999}

\maketitle

\label{firstpage}

\begin{abstract}
We explore the wind-driving mechanism of giant stars through the nearby (117 pc), intermediate-luminosity ($L \approx 1600$ L$_\odot$) star EU Del (HIP 101810, HD 196610). APEX observations of the CO (3--2) and (2--1) transitions are used to derive a wind velocity of 9.51 $\pm$ 0.02 km s$^{-1}$, a $^{12}$C/$^{13}$C ratio of 14 $^{+9}_{-4}$, and a mass-loss rate of a few $\times$ 10$^{-8}$ M$_\odot$ yr$^{-1}$. From published spectra, we estimate that the star has a metallicity of [Fe/H] = --0.27 $\pm$ $\sim$0.30 dex. The star's dusty envelope lacks a clear 10-$\mu$m silicate feature, despite the star's oxygen-rich nature. Radiative transfer modelling cannot fit a wind acceleration model which relies solely on radiation pressure on condensing dust. We compare our results to VY Leo (HIP 53449), a star with similar temperature and luminosity, but different pulsation properties. We suggest the much stronger mass loss from EU Del may be driven by long-period stellar pulsations, due to its potentially lower mass. We explore the implications for the mass-loss rate and wind velocities of other stars.
\end{abstract}

\begin{keywords}
stars: mass-loss --- circumstellar matter --- stars: winds, outflows --- stars: AGB and post-AGB --- stars: variables --- stars: individual: EU Del
\end{keywords}


\section{Introduction}
\label{IntroSect}

The wind-driving mechanism of giant stars is poorly determined \citep[e.g.][]{Woitke06b}. Conventionally, low- and intermediate-mass stars ($\approx$0.8--8 M$_\odot$) ascending the asymptotic giant branch (AGB) are thought to first exhibit a magneto-acoustically driven wind, which later transitions to a pulsation-enhanced, dust-driven wind (e.g.\ \citealt{DHA84,WlBJ+00}). Initially, magneto-acoustic waves propagating through the outer stellar atmosphere heat a stellar chromosphere providing energy for an outflow, either directly or via solar-like magnetic reconnection events. Later, $\kappa$-mechanism pulsations are stochastically excited by the star's large atmospheric convective cells. These pulsations levitate its outer atmosphere. The levitated material cools and condenses into dust. The opaque dust absorbs outgoing stellar radiation and re-radiates it isotropically, imparting net outward momentum, which forces it from the star. Collisional coupling with gas particles ejects the entire envelope into the interstellar medium \citep[e.g.][]{Habing96}.

The mechanism by which the wind is ejected is important not only for modelling this stage of evolution, but also for its extrapolation outside well-measured environments. It is particularly important in metal-poor stars, hence the early Universe. Here, stars of a given mass are smaller and hotter, so a pulsation-driven wind must impart more energy to levitate material to the same dust-formation radius. Less refractory material means a lower dust-to-gas ratio, decreasing the efficacy and velocity of a dust-driven wind. Separation of gas and dust can occur, leading to a drift velocity, preventing the escape of the gas, and perhaps depleting the star of dust-forming elements. 

Despite this, prolific dust condensation is seen around metal-poor stars, indicating these stars somehow drive substantial stellar winds \citep{SMM+10,MvLS+11}. Dust production begins at $\sim$600--2000 L$_\odot$ \citep{MBvLZ11,MSS+12,MZB12}. However, there is insufficient opacity to drive the wind, particularly among oxygen-rich stars, until the star reaches higher luminosities \citep{GWG+10}. By measuring the wind expansion velocity and total mass-loss rate of stars as a function of properties such as luminosity and metallicity, we can determine the point at which the transition to a pulsation and/or dust-driven wind occurs.

Such measurements are usually derived from (sub-)mm CO lines. However, most metal-poor stars with well-known properties lie at large distances, in globular clusters or dwarf galaxies. Their distance and interstellar radiation environment make them challenging to observe. A search for closer dust-producing giant stars with \emph{Hipparcos} parallaxes identified several nearby targets \citep{MZB12}.

\begin{figure}
\centerline{\includegraphics[height=0.47\textwidth,angle=-90]{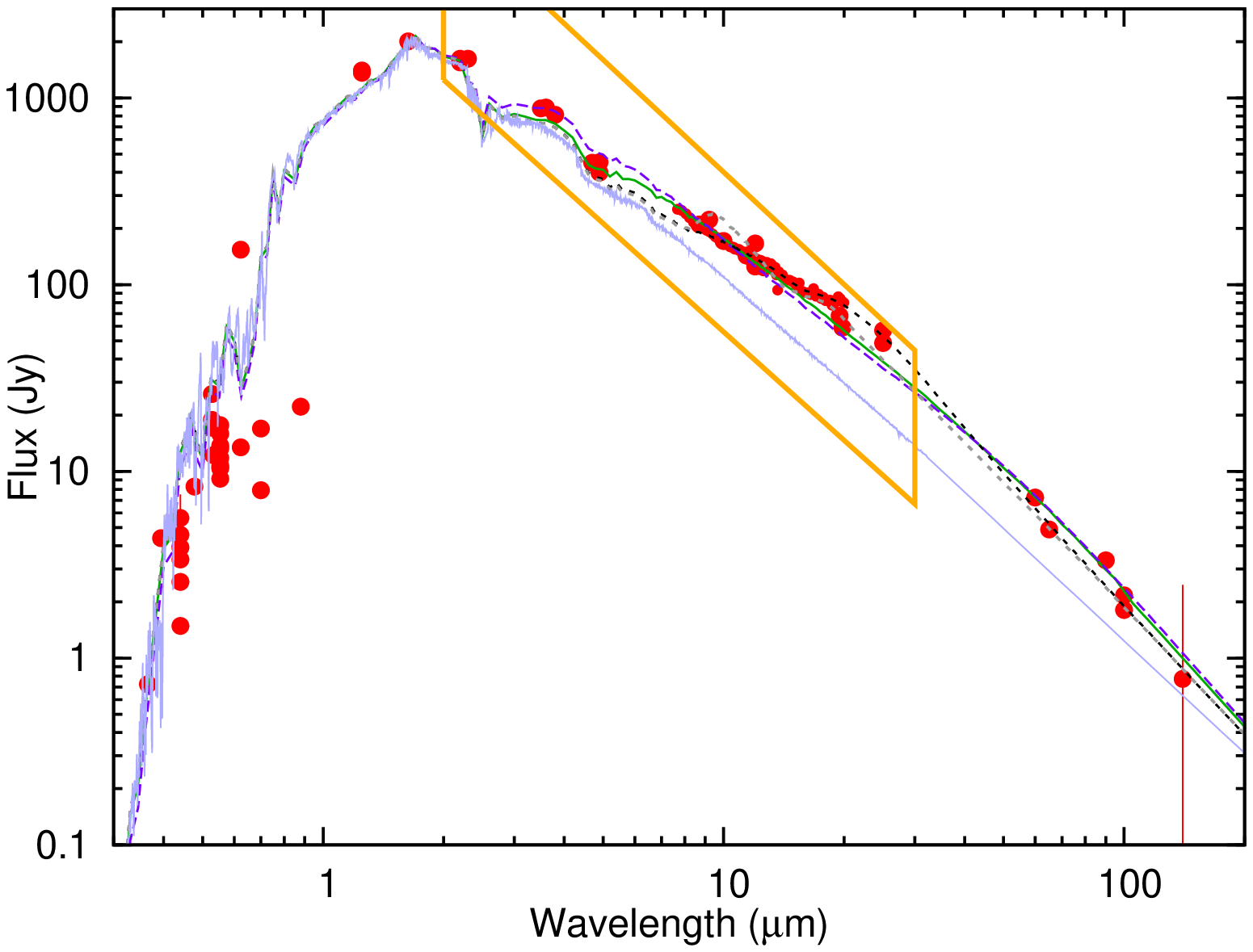}}
\centerline{\includegraphics[height=0.47\textwidth,angle=-90]{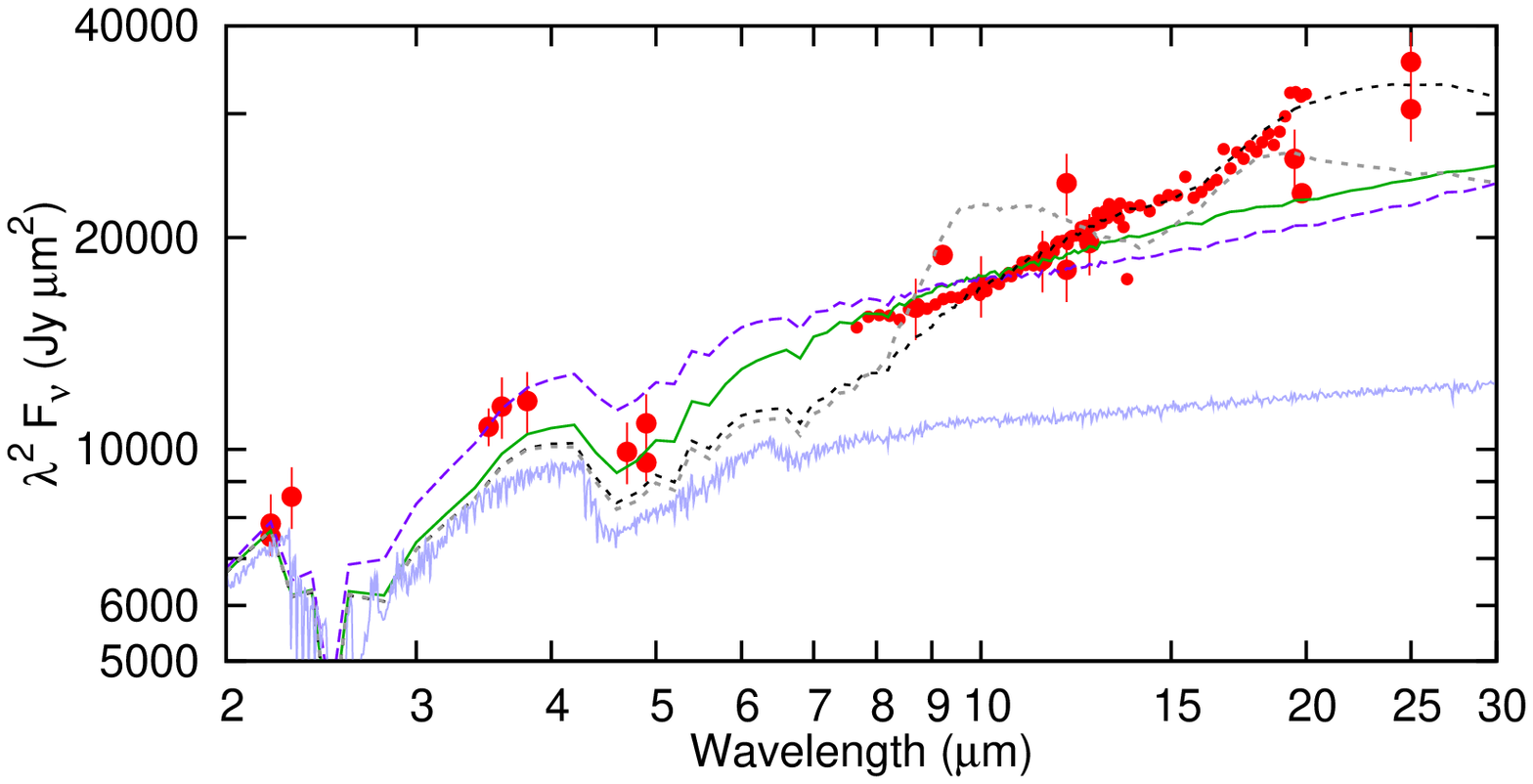}}
\centerline{\includegraphics[height=0.47\textwidth,angle=-90]{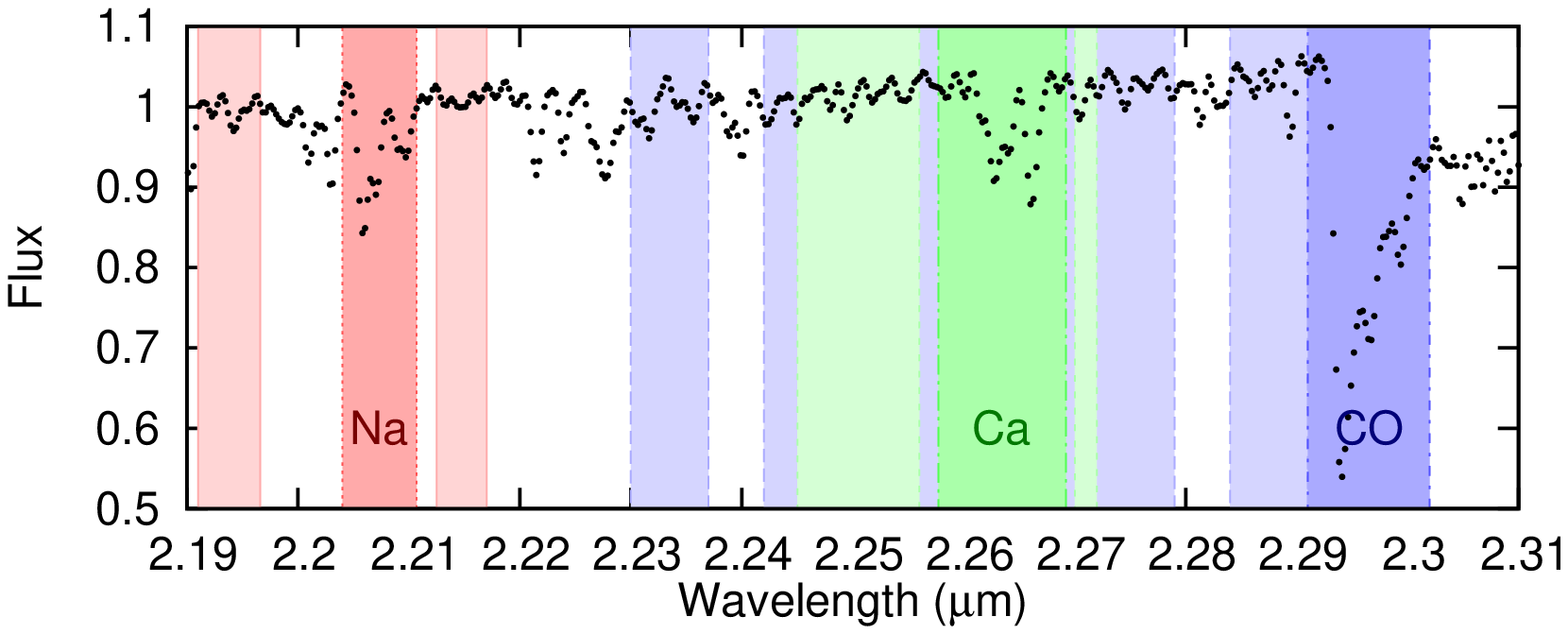}}
\caption{Top panel: Spectral energy distribution of EU Del (large red points) including the \emph{IRAS}-LRS spectrum (small red points), with a stellar atmosphere model (thin blue line) and {\sc dusty} model fits (thicker green, purple, grey and black lines; see Section \ref{RadWind}). The very thick orange parallelogram shows the portion expanded in the middle panel. See text for details of the models. Middle panel: the infrared portion of the spectrum, shown in Rayleigh-Jeans units to emphasise dust features. A naked stellar photosphere (thin blue line) should be roughly flat. A reddened spectrum indicates dust. Bottom panel: the $K$-band spectrum \citep{WMHE00} used for the metallicity determination, with line/continuum regions (dark/light boxes) marked for each line.}
\label{SEDFig}
\end{figure}

One of these targets was EU Del (HIP 101810, HD 196610), the stellar parameters of which are listed in Table \ref{StarTable}. EU Del is among the closest bright giant stars. Table \ref{StarTable} lists its properties and Figure \ref{SEDFig} shows its spectral energy distribution (SED). It is a semi-regular variable, with a characteristic period of $\sim$60 days \citep{SDZ+06}. \citet{WSP+11} estimated that EU Del had a metallicity of [Fe/H] $\sim$ --1. However, their metallicity was poorly constrained and based on an automated fitting procedure that was not optimised for cool stars. An \emph{IRAS} LRS spectrum of the star originally led it to be classified as `naked' (i.e.\ not producing dust; \citealt{SP95}). However, detailed modelling of the entire spectral energy distribution shows it to have considerable infrared excess, indicative of circumstellar dust \citep{MZB12}. Unusually for an oxygen-rich star, it does not show the typical 10-$\mu$m silicate or 11-$\mu$m alumina dust features, which partly explains the difference in classifications\footnote{\citealt{SP95} detect a dust emission by determining a semi-empirical function to the stellar continuum between 7.67 and 14.03 $\mu$m. This is divided out and the normalised spectrum integrated. Stars with $<$8 per cent emission are classed as naked. Thus stars with a low-contrast infrared excess but no dust features in their spectra can still be classed as naked.}. This is typical of metal-poor stars in globular clusters \citep{MSZ+10}. In this paper, we improve the determination of EU Del's metallicity and present new observations of the star's CO lines, placing its circumstellar wind in context with other mass-losing giant stars.


\section{Stellar parameters}

\begin{center}
\begin{table}
\caption{Properties of EU Del.}
\label{StarTable}
\begin{tabular}{l@{}c@{}r}
    \hline \hline
Parameter  & Value  & Ref. \\
    \hline
RA			& 20 37 54.728				& 1 \\
Dec			& +18 16 06.89				& 1 \\
Distance ($d$)		& 117 $\pm$ 7 pc			& 1 \\
RA proper motion	& +35.44 mas yr$^{-1}$			& 1 \\
Dec proper motion	& +61.57 mas yr$^{-1}$			& 1 \\
Tangential velocity	& 38 km s$^{-1}$			&   \\
Radial velocity		& --66 km s$^{-1}$			& 2 \\
Spectral type		& M6					& 3 \\
Effective temperature
\rlap{($T_{\rm eff}$)}	& 3100 K				& 4 \\
\ 			& 3243 K				& 5 \\
\ 			& 3227 K				& 6 \\
\ 			& 3508 K				& 7 \\
Luminosity ($L$)	& 1585 L$_\odot$			& 6 \\
log($g$)		& 0.39					& 4 \\
Radius ($R$)		& 127.5 R$_\odot$			& 6 \\
Assumed mass ($M$)	& $\approx$0.6 M$_\odot$		& \\
Assumed initial mass	& $\sim$0.9 M$_\odot$			& \\
Escape velocity for $M$	& 42 km s$^{-1}$			& \\
Variability type	& SRV (O-rich)				& 8 \\
Period			& 59.7 days				& 9 \\
\ 			& 60.8, 132.6, 235.3, 67.3, 44.0 days	& 10 \\
Amplitude (optical)	& 149, 113, 91, 75, 61 mmag		& 10 \\
\protect{[Fe/H]}	& $\sim$--1				& 4 \\
\ 			& --0.27 $\pm$ 0.30			& 11 \\
$^{12}$C/$^{13}$C	& 14$^{+9}_{-4}$			& 11 \\
    \hline
\multicolumn{3}{p{0.45\textwidth}}{References: (1) \citet{vanLeeuwen07}; (2) \citet{dBE12}; (3) \citet{KM89}; (4) \citet{WSP+11}; (5) \citet{CPI+13}; (6) \citet{MZB12}; (7) \citet{SlCCdSC10}; (8) \citet{MMAL01}; (9) \citet{SDZ+06}; (10) \citet{TBK+09}; (11) this work.}\\
    \hline
\end{tabular}
\end{table}
\end{center}

\subsection{Kinematic origin of EU Del}

\citet{AF12} provide galactocentric velocity components for EU Del of $U$ = --64.3, $V$ = --37.8 and $W$ = +18.9 km s$^{-1}$. They calculate that this velocity gives it an eccentric orbit, ranging in galactocentric radius from 4.5 to 8.4 kpc, implying that EU Del is close to apocentre. Using the Galactic Orbit Calculator\footnote{http://www.astro.utu.fi/galorb.html}, we find that the orbit of EU Del extends out of the plane by up to 300 pc. Thus, the star is normally around 150 pc from the plane.

Although the rotational velocity component ($V$) is large for a thin disk star \citep{IBJ12}, the extension from the plane is fairly small. The scale height of the thin disk and thick disks are $\sim$300 and $\sim$1350 pc, respectively, with the thick disk providing $\sim$2 per cent of the stars in the solar neighbourhood \citep{GR83}. This translates to $\sim$3 per cent of stars at 150 pc from the plane, thus EU Del is most likely a thin disk star.


\subsection{Metallicity of EU Del}
\label{MetalSect}

\subsubsection{Spectral discriminants}

Strong molecular bands dominate the optical spectrum of EU Del \citep{VGR+04}. These swamp the atomic lines normally used to determine stellar abundances. Near-IR spectra of EU Del \citep[][for the $J$, $H$ and $K_{\rm s}$ bands, respectively]{WH97,MEHS98,WMHE00} are not of sufficiently high resolution to enable detailed abundance measurements.

Instead, we compute [Fe/H] from the $K$-band spectrum (\citealt{WMHE00}; Figure \ref{SEDFig}), using the windowing functions of \citet{RSFD00} and \citet{FSRD01}. \citet{RSFD00} provide two methods to establish [Fe/H], based on the equivalent widths of the 2.21-$\mu$m Na doublet and 2.26-$\mu$m Ca triplet lines, and the 2.29-$\mu$m CO $\nu$=2$\rightarrow$0 bandhead. Both methods fit empirical second-order polynomials to spectra of 77 globular cluster giant stars, which have similar effective temperatures to our target. Method 1 is based only on the equivalent widths of the above spectral features. Method 2 also fits the $(J-K_s)$ colour and uses the $M_{Ks}$ absolute magnitude of the star.

We use Equations 3 \& 4 of \citet{FSRD01}, which encode these two methods, to determine [Fe/H]. We measure equivalent widths of 3.53 $\pm$ 0.18, 2.63 $\pm$ 0.38 and 22.68 $\pm$ 0.53 \AA\ for the Na, Ca and CO features, respectively. Methods 1 and 2 give [Fe/H] = --0.30 $\pm$ 0.05 and --0.23 $\pm$ 0.06 dex, respectively.

The quoted errors are almost certainly underestimates, as no uncertainties are given on the fitting factors in either method. Based on the scatter of the [Fe/H] of individual stars from the `known' cluster average \citep[][their figures 11 \& 12]{FSRD01}, we can expect the typical uncertainty to be $\sim\pm$0.2 dex. 

We also note that no iron lines are used, thus the determination of [Fe/H] assumes a particular stellar composition. We must therefore account for the fact that the formalism is calibrated on $\alpha$-enhanced stars in globular clusters. These stars are typically Ca-rich and C-poor, which have opposing effects on the abundance determination (Na abundance can vary; \citealt{RCGS13}). Galactic field stars at [Fe/H] $\approx$ --0.3 dex have only slightly positive [Na/Fe] and [Ca/Fe] ratios \citep{EAG+93}, which we estimate adds a further $\sim\pm$0.1 dex uncertainty onto the metallicity estimate. Taking these uncertainties and averaging between the two methods, we find [Fe/H] = --0.27 $\pm$ $\sim$0.30 dex.

\subsubsection{Photometric discriminants}

The position of EU Del in the Hertzsprung--Russell diagram \citep{MZB12} is not significantly to the blue of the main red giant branch, as would be expected for a markedly metal-poor star (e.g. \citealt{MGB+08}). However, interstellar reddening has not been taken into account. For giant stars, reddening of $E(B-V) = 0.01$ and $R_V = 3.1$ decreases the fitted effective temperature by $\approx$6 K \citep{MvLD+09}. The reddening towards EU Del is hard to compute as it is embedded within the Galaxy's thin disc. We have used the extinction maps of \citet{LVV+14} to estimate that $E(B-V) \approx 0.015$, so it can be considered negligible.

EU Del is therefore most likely slightly metal-poor, but given the sizable and poorly quantified uncertainties on this measurement, we cannot reliably distinguish its metallicity from solar abundances. However, this value improves on existing determinations and contradicts the [Fe/H] = --1 value of \citet{WSP+11}.



\subsection{Wind velocity of EU Del}
\label{VexpSect}

\begin{center}
\begin{table*}
\caption{Wind parameters of EU Del from APEX observations.}
\label{WindTable}
\begin{tabular}{lcccccc}
    \hline \hline
Parameter  & Unit        & $^{12}$CO $J$=2--1 & $^{12}$CO $J$=3--2 & $^{13}$CO $J$=3--2 & Global \\
    \hline
Frequency                     & GHz                 & 230.538                   & 345.796                   & 330.587                & \nodata \\
Resolution                    & km s$^{-1}$         & 0.099                     & 0.066                     & 0.069                  & \nodata \\
Noise                         & K                   & 0.023                     & 0.062                     & 0.022                  & \nodata \\
Intensity                     & K km s$^{-1}$       & 1.96 $\pm$ 0.04           & 3.57 $\pm$ 0.08           & 0.26 $\pm$ 0.12        & \nodata \\
Amplitude                     & K                   & 0.1031 $\pm$ 0.0017       & 0.187 $\pm$ 0.004         & 0.013 $\pm$ 0.005      & \nodata \\
\                             & Jy                  & 4.02 $\pm$ 0.07           & 7.69 $\pm$ 0.16           & 0.54 $\pm$ 0.21        & \nodata \\
Half-width                    & km s$^{-1}$         & 9.492 $\pm$ 0.015         & 9.522 $\pm$ 0.019         & 9.86 $\pm$ 0.65        & 9.505 $\pm$ 0.021\\
Centre                        & km s$^{-1}$         & --51.93 $\pm$ 0.04        & --51.77 $\pm$ 0.04        & --51.94 $\pm$ 0.60     & --51.85 $\pm$ 0.10\\
Boxcar $\chi^2_{\rm red}$     & \nodata             & 1.087                     & 1.322                     & 1.224                  & \nodata \\
$\dot{M}_{\rm Ramstedt}$      & M$_\odot$ yr$^{-1}$ & 5.1 $^{+7.5}_{-3.8}$ $\times$ 10$^{-8}$ & 3.4 $^{+7.8}_{-3.2}$ $\times$ 10$^{-8}$ & \nodata                & 4.7 $^{+5.3}_{-3.7}$ $\times$ 10$^{-8}$\\
    \hline
\multicolumn{6}{p{0.95\textwidth}}{Velocities are in the $v_{\rm LSR}$ frame, temperatures are antenna temperatures. The global mass-loss rate was produced by the logarithmic averaging of the $^{12}$CO $J$=3--2 and 2--1 lines.}\\
    \hline
\end{tabular}
\end{table*}
\end{center}

\begin{figure}
\centerline{\includegraphics[height=0.47\textwidth,angle=-90]{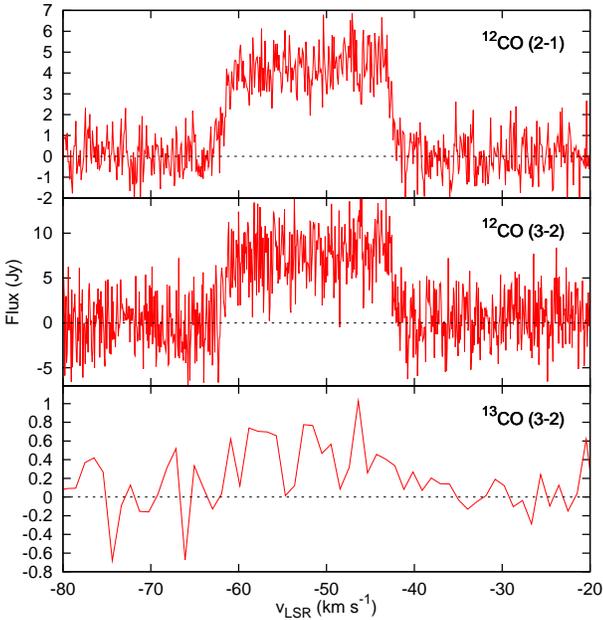}}
\caption{APEX spectra of EU Del. The $^{13}$CO spectrum has been smooth to aid visibility.}
\label{SpecFig}
\end{figure}


We obtained 230 GHz (850 $\mu$m) and 345 GHz (450 $\mu$m) observations of EU Del from the Swedish Heterodyne Facility Instrument (SHeFI) on the Atacama Pathfinder Experiment (APEX) telescope, on 2013 Aug 19 and 2014 Aug 18, respectively. We reduced the data using {\sc class} and {\sc xs}\footnote{XS is a software package developed by P. Bergman to reduce and analyse single-dish spectra. It is publicly available from: ftp:// yggdrasil.oso.chalmers.se.}. Antenna temperature data were baseline subtracted by removing a straight line, fitted to the baseline flux. Temperatures were then converted to physical units using 39 Jy K$^{-1}$ at 230 GHz and 41 Jy K$^{-1}$ at 345 GHz\footnote{http://www.apex-telescope.org/telescope/efficiency/}. Table \ref{WindTable} lists the resulting wind parameters.

These observations covered the $^{12}$CO $J$=2--1, $^{12}$CO $J$=3--2 and $^{13}$CO $J$=3--2 transitions. Figure \ref{SpecFig} shows the spectra for these lines. The $^{12}$CO lines shows rectangular profiles, consistent with an unresolved, optically thin line (the $^{13}$CO line lacks the clarity needed to determine the shape accurately). A top-hat function was fit to each line using $\chi^2$ minimisation, with free parameters in expansion velocity (half-width), amplitude and central velocity. The best-fit parameters were passed to the {\sc idl} optimization routine, {\sc mpfit} \citep{Markwardt09}, which uses the Levenberg--Marquardt technique to fit a least-squares, best-fit model. {\sc Mpfit} re-fitted the top-hat function (including a baseline offset) and computed the errors on the model parameters from the covariance matrix. From this, we derive the wind parameters listed in Table \ref{WindTable}.


\subsection{Mass-loss rate}
\label{MLRSect}

An approximate mass-loss rate estimate can be obtained by using scaling relations derived from known stars. However, EU Del is relatively low luminosity and (perhaps) metal-poor compared to the stars for which these relations are calibrated. Such scaling relations are therefore poorly constrained in this regime. We adopt the relation of \citet[][their equation A.1]{RSOL08} to produce the mass-loss rates in Table \ref{WindTable}. To compute these, we have assumeed beam sizes of 26.6$^{\prime\prime}$ at 230 GHz and 17.3$^{\prime\prime}$ at 345 GHz. We also assumed that Ramstedt's CO fraction (1/5\,000th of the wind at solar metallicity) decreases linearly with iron abundance. The formal errors are approximate, and the mass-loss rate can only be constrained to be between $\sim$10$^{-8}$ and $\sim$10$^{-7}$ M$_\odot$ yr$^{-1}$, thus we quote the mass-loss rate in the discussion as `a few' $\times$ 10$^{-8}$ M$_\odot$ yr$^{-1}$.





\subsection{$^{12}$C/$^{13}$C ratio}
\label{C12C13Sect}

The ratio of the $^{12}$CO (3--2) and $^{13}$CO (3--2) lines provides a $^{12}$C/$^{13}$C ratio for EU Del of 14$^{+9}_{-4}$, where the large uncertainty arising from the weak detection of the $^{13}$CO line. This ratio is typical for a star which has experienced first dredge-up but not third dredge-up (A.~Karakas, private communication; see also, e.g., \citealt{RL05}). It also matches the near-equilibrium values observed in metal-poor RGB-tip field stars \citep{GSCB00}. However, the accuracy of this measurement depends on several assumptions:
\begin{itemize}
\item {\it The emission is optically thin.} This is attested by the flatness of the $^{12}$CO line peaks, which will be more opaque than the $^{13}$CO lines. An optically thick line would have produced a parabolic shape to the line.
\item {\it The $^{12}$CO and $^{13}CO$ are emitted from the same region.} The optical thinness of the envelopes and similarity in the energy levels of $^{12}$CO $J$=3 and $^{13}$CO $J$=3 transitions (33.19 and 31.73 K, respectively) suggest the emission should come from very similar regions of the envelope.
\item {\it The $^{12}$CO/$^{13}$CO ratio is equal to the $^{12}$C/$^{13}$C ratio.} Selective photo-dissociation of $^{13}$CO can occur if it is optically thin to UV radiation ($>$11.1 eV) and $^{12}$CO is optically thick. Chemical fractionation can allow $^{13}$CO to form preferentially, thus providing the opposite effect (e.g., \citep{MGH88,VML13,RO14}. This can provide an additional uncertainty of typically several tens of percent in the $^{12}$C/$^{13}$C ratio. This uncertainty is not included here, as the wind of EU Del is optically thin in the $^{12}$CO transitions.
\end{itemize}


\section{The wind driving mechanisms}
\label{DrivingSect}

In this section, we explore the ability of three mechanisms to drive the wind: radiation pressure on dust, energy transfer from pulsations, and release of magnetic energy from the stellar surface. In order to reproduce the minimum mass-loss rate and wind velocity listed in Table \ref{WindTable}, a kinetically-driven wind requires $\dot{E} = \dot{M}v^2/2 \gtrsim 2.7 \times  10^{22}$ W of kinetic energy, plus $\gtrsim$5.3 $\times$ 10$^{23}$ W of energy to overcome the gravitational potential ($\dot{E} = -G M \dot{M} / R$), or $\gtrsim$5.6 $\times$ 10$^{23}$ W in total.




\subsection{A radiation-driven wind}
\label{RadWind}

In a radiation-driven wind, momentum from outgoing stellar photons is transferred to the dust grains (see Section \ref{IntroSect}). The densest, fastest wind that can be accelerated by this mechanism is:
\begin{equation}
\frac{L}{c} \gtrsim \dot{M} v ,
\end{equation}
where the luminosity $L = 1585$ L$_\odot$ for EU Del \citep{MZB12}. For a 9.5 km s$^{-1}$ wind, $\dot{M} \leq 3.4 \times 10^{-6}$ M$_\odot$ yr$^{-1}$. However, this limit does not take the stellar gravity into account. Escape velocity at the canonical dust-formation radius ($\approx$2 $R_\ast$) is $\sim$21 km s$^{-1}$, which reduces the maximum mass-loss rate to $\dot{M} \lesssim 1.0 \times 10^{-6}$ M$_\odot$ yr$^{-1}$. The exact value depends on the actual dust-formation radius and the momentum already imparted to the escaping material by stellar pulsations.

In addition to these factors, only a fraction of the stellar light is reprocessed. \citet{MZB12} presented an SED compiled from literature photometry and spectroscopy for EU Del, showing the star to have a substantial infrared excess, with $\sim$1.4 per cent of light being reprocessed \citep{MZB12}. This reduces the upper limits such that $\dot{M} \lesssim 1.5 \times 10^{-8}$ M$_\odot$ yr$^{-1}$ for a radiation-driven wind. The range of possible mass-loss rates is $\sim$10$^{-8}$ to 10$^{-7}$ (we remind the reader that formal errors are poorly determined). There is little overlap between the range of possible mass-loss rates from observations and the range that be sustained from radiation pressure on dust.

This $\sim$1.4 per cent can be increased if photons are scattered by large dust grains, rather than absorbed, as scattering does not change the overall shape of the SED \citep{Hoefner08}. To better address radiation transport within the wind, we have modelled the SED of EU Del using the {\sc dusty} radiative transfer code \citep{IE97}. {\sc Dusty}'s {\tt density-type = 3} setting computes a radiation-driven wind structure by solving the wind hydrodynamics, including dust drift, and computes limits where the gravitational attraction from the host star becomes important. We use a {\sc bt-settl} model atmosphere at 3300 K, log($g$) = 0, [Fe/H] = --0.5 \citep{AGL+03} to represent the underlying star (Figure \ref{SEDFig}).

A radiative transfer dust model requires knowledge of the dust species involved. The infrared spectrum of EU Del shows very weak dust features. A weak feature may exist near 13 $\mu$m due to alumina \citep{SKGP03} and a 20-$\mu$m silicate feature is suggested by the rising spectrum between 15 and 20 $\mu$m, but it is difficult to determine whether these are real. Weak or absent features are typical of metal-poor stars in globular clusters \citep{MvLD+09,SMM+10,MBvL+11,MvLS+11}. In globular clusters, metallic iron has been suggested to be the dominant component of such stars \citep{MSZ+10}, but carbonaceous grains remain an alternative possibility \citep{HA07}. Since we cannot readily identify the dust species around EU Del, we explore several different options.

We have created four different {\sc dusty} models. For all models, we assume a standard Mathis--Rumpl--Nordsiek size distribution \citep{MRN77}, i.e.\ that the number ($n$) of grains with size $a$ is given by $n(a) \propto a^{-q}$ where $q$ is typically 3.5. Typically, sizes are limited between 0.025 and 0.25 $\mu$m for silicate grains. {\sc Dusty} uses Mie theory to calculate the appropriate opacity. For all models, we assume that the dust:gas ratio is $\sim$1:372, based on scaling the solar-metallicity value of $\sim$1:200 to the [Fe/H] $\sim$ --0.27 metallicity derived above. Our modifications to this model are as follows:
\begin{itemize}
\item Pure metallic iron, using optical constants from \citet{OBA+88}, with grain size limits 0.005 to 0.03 $\mu$m (green, solid line in Figure \ref{SEDFig}). A condensation temperature of 800 K is assumed. This reproduces the spectrum between 4 and 10 $\mu$m well, but fails to capture the excess emission between 10 and 25 $\mu$m. The small grain size is needed to reduce the 100-$\mu$m emission.
\item Pure amorphous carbon, using optical constants from \citet{Hanner88}, with the same grain size (purple, dashed line in Figure \ref{SEDFig}). A condensation temperature of 800 K is assumed. This fits the data between 3 and 4 $\mu$m better, but overestimates the 4--5 $\mu$m contribution and underestimates the 10-25 $\mu$m flux even more than the metallic iron model.
\item A mix of amorphous silicates (80 per cent; \citealt{DL84}) and alumina (20 per cent; \citealt{BDH+97}) with large grains. Grain sizes are limited to 0.005 to 1 $\mu$m, and a condensation temperature of 1000 K is assumed (grey, dotted line in Figure \ref{SEDFig}). This underestimates the flux between 2 and 10 $\mu$m and at 25 $\mu$m and produces an unobserved 10-$\mu$m peak.
\item The same mix, but with a single population of 3-$\mu$m grains (black, dotted line in Figure \ref{SEDFig}). This under-estimates the flux between 2 and 10 $\mu$m, but reproduces the SED at all longer wavelengths.
\end{itemize}
The mixture of silicates to alumina in the final two models were varied, with the above 80:20 mixture best fitting the observed IR spectrum. A higher alumina ratio produces an unobserved 11-$\mu$m peak in preference to a 10-$\mu$m peak.

The fourth model is conceptually difficult to form, as it requires large silicate grains without any smaller silicate grains to form them from. We consider this model unlikely, but it may be physically possible if small alumina grains are coated with silicates. It seems likely that iron (or a similar species) contriubtes to the spectrum, in order to reproduce the 2--10 $\mu$m dust emission. Iron cannot condense near the star due to its high opacity \citep{BH12}. A gradient of increasing iron content in silicate grains at larger radii from the star may be possible \citep{BHAE15}. This would hold the grain temperature near the condensation temperature ($\sim$1000 K), leading to a greater 2--10 $\mu$m emissivity than we model here. Some combination of all of these factors may lead to layered grains more complex than the grain population we model here.

The mass-loss rates derived from our fits are $\dot{M} = 5.7, 1.3, 1.3$ and $1.0 \times 10^{-7}$ M$_\odot$ yr$^{-1}$, respectively. The latter three are similar to the CO mass-loss rate of $\sim$10$^{-8}$ to 10$^{-7}$ M$_\odot$ yr$^{-1}$. Metallic iron produces a mass-loss rate greater than that measured (Section \ref{MLRSect}), however we argue in the previous paragraph why this does not necessarily mean it is invalid.

Crucially, however, the expansion velocities in all cases are very small: only 2.1, 2.4, 7.5 and 5.5 km s$^{-1}$, respectively. The metallic iron and amorphous carbon velocities are below the velocities at which the {\sc dusty} models can be considered valid (5 km s$^{-1}$) and far below the expansion velocity of the wind of EU Del (9.5 km s$^{-1}$). Any increase in the wind velocity would require increasing the mass-loss rate proportionally and altering the velocity structure. The third (large-grained silicate--alumina conglomerate) model could achieve a 9.5 km s$^{-1}$ wind velocity if the star were of solar metallicity, however it fails to reproduce the observed spectrum. The fourth (3-$\mu$m silicate) model reaches 7.5 km s$^{-1}$ at solar metallicity but, as noted, is physically difficult to create. More importantly, {\sc dusty} does not account for the decrease in effectiveness due to stellar gravity. The minimum applicable mass (``$<M$'') warning in the {\sc dusty} software indicates that gravity has a substantial unmodelled effect on these models. We have explored a number of other options to try to reproduce the mass-loss rate and wind velocity observed without success.

We therefore conclude that EU Del is unlikely to have a primarily radiation-driven wind, although the velocity may be enhanced by radiation pressure on dust. Should a different energy source provide most of the power for the wind, the dust particles will not start from a resting levitated layer above the star. They will therefore spend less time in the radiation-driving zone. Hence, the time-integrated acceleration the wind experiences due to radiation pressure will be less than suggested by the velocities quoted above.


%



\subsection{A magnetically-driven wind}
\label{MagWind}

Without direct observation of the magnetic field strength and the stellar rotation period, it is difficult to reasonably determine the energy in a magnetically-driven wind. Magnetic fields of luminous ($\sim$10$^3$ L$_\odot$) giants tend to be weak ($<$1 G; \citealt{AKAC+15}), although a few individual stars have a field of a few Gauss \citep[e.g.][]{LAF+14,SWL15}. Rotation periods are sufficiently slow that the lines are not observably broadened.

An alternative approach is to compare to scaling relations based on empirical mass-loss rates. Relations such as those of \citet{Reimers75} and \citet{SC05} accurately represent mass-loss rates of stars across a large part of the Hertzsprung--Russell diagram where magnetic activity is thought to drive the stellar wind. These relations prescribe mass-loss rates (in M$_\odot$ yr$^{-1}$) of:
\begin{eqnarray}
\dot{M}_{\rm Reimers} &\!\!\!\!\!\!=\!\!\!\!\!\!& 4 \times 10^{-13} \eta_{\rm Reimers} \frac{LR}{M} \\
{\rm \ and} && \nonumber\\
\dot{M}_{\rm SC} &\!\!\!\!\!\!=\!\!\!\!\!\!& 4 \times 10^{-13} \eta_{\rm SC} \frac{LR}{M} \left(\frac{T_{\rm eff}}{4000 {\rm K}}\right)^{3.5} \left(1+\frac{g_{\odot}}{4300g}\right) , \nonumber\\
\end{eqnarray}
where $\eta_{\rm Reimers}$ and $\eta_{\rm SC}$ are fitting parameters.	

The value of $\eta$ may depend on the environment under investigation and is likely a function of more parameters \citep[e.g.][]{SC05}. In globular clusters, it is well calibrated on the red giant branch (RGB), at $\eta_{\rm Reimers} = 0.477 \pm 0.132$ and $\eta_{\rm SC} = 0.172 \pm 0.047$ \citep[][including systematic errors]{MZ15b}. \emph{Kepler} measurement of stars in the open cluster NGC 6791 suggests $0.1 \lesssim \eta_{\rm Reimers} \lesssim 0.3$ \citep{MBS+12}\footnote{However, it is likely that $\eta_{\rm Reimers} \sim 0.1$ would be sufficiently low to delay most mass loss to the dust-enshrouded `superwind' phase of the AGB. This enhanced `superwind' is not seen in NGC 6791 \citep{vLBM08}}. $L$ and $R$ take the values from Table \ref{StarTable}, which yields $log(g) = 0.00$.

Adopting these values gives $\dot{M}_{\rm Reimers} = 6 \times 10^{-8}$ M$_\odot$ yr$^{-1}$ and $\dot{M}_{\rm SC} = 8 \times 10^{-8}$ M$_\odot$ yr$^{-1}$. These are within the range of mass-loss rates obtained from the CO lines, implying that magnetic energy could drive the wind. Whether it does so depends on whether these empirical laws accurately describe the transfer of magnetic energy to the wind of bright giant stars, which is uncertain.

Qualitatively, no indicators of a magnetically driven wind are present in the spectrum. Conventionally, magnetic heating of plasma above the surface produces a chromosphere (but not a full corona; \citealt{HM80}). This chromosphere has a low filling factor, but exhibits emission components to chromospherically active lines (e.g.\ \citealt{DHS+94,MvL07,MAD09}). Such lines include H$\alpha$, H$\beta$, H$\gamma$ (etc.), the infrared calcium triplet, the Mg II K and Na II D lines. Emission components to these lines are variable in strength, but rarely disappear completely. No emission components are found in the optical spectrum of \citet{VGR+04}.

Emission components are normally observed to disappear in more luminous stars, probably when the chromosphere becomes destabilised by the stellar pulsation. They later give way to stronger, less structured emission lines as atmospheric shocks from the puslations again heat the circumstellar environment \citep{DHA84,MvL07}. In conclusion, we cannot rule out a magnetically driven wind, but neither does it appear likely from qualitative indicators.




\subsection{A pulsation-driven wind?}

Over the course of the pulsation cycle, a fraction of the radiation from the stellar interior is trapped during the rising phase of the stellar pulsation and released during the falling phase. In a pulsation-driven wind, the kinetic energy given to the wind must be less than half the energy trapped by the stellar atmosphere during the pulsation cycle. Hence pulsations have sufficient energy to drive a wind if:
\begin{equation}
\frac{\dot{M}v^2}{L} < \frac{1}{2} \Delta ,
\label{PulsEq}
\end{equation}
where $\Delta$ is the peak-to-peak pulsation amplitude, measured as a fraction of the star's bolometric flux \citep{vanLoon02,vLCO+08}. This should be very similar to the light curve amplitude at the peak of the star's SED. Light curves from the Diffuse Infrared Background Experiment (DIRBE), although of poor signal-to-noise \citep{PSK+10}, show the amplitude of EU Del rises from $\sim$0.05 mag at 4.9 $\mu$m to $\sim$0.1 mag at 2.2 $\mu$m (see also \citealt{TBK+09}). The amplitude increases further to $\sim$0.45 mag in the \emph{Hipparcos} data at 0.53 $\mu$m \citep{vanLeeuwen07}. The star's SED peak is in the $H$-band. The $H$-band photometric amplitude is likely to be slightly larger than that at 2.2 $\mu$m, but smaller than the optical variation. Assuming the phase lag between the optical and infrared lightcurves is small, we can conservatively expect that the bolometric variability ($\Delta$), is $\sim$10 per cent of the total intensity.

Half of this is available to drive the wind, or 1.2 $\times$ 10$^{29}$ W. Were this energy liberated into the wind, it could support a mass-loss rate of a staggering 0.021 M$_\odot$ yr$^{-1}$. As highlighted by \cite{vLCO+08}, only a tiny fraction (one part in $\sim$10$^6$) of this needs to be liberated to drive the wind.

A more realistic estimate may come from pulsation growth rates. The dimensionless growth rate ($\dot{A}$) defines the fractional increase in pulsation amplitude per pulsation cycle. Since most stars have stable pulsation amplitudes, this defines the maximum fraction of pulsation energy that could be expended to drive material from the star (the remainder of which will likely heat the outer atmosphere), thus:
\begin{equation}
\frac{\dot{M}v^2}{L} < \frac{1}{2} \dot{A} \Delta .
\label{PulsEq}
\end{equation}
For EU Del, we can expect growth rates of $\sim$10 per cent \citep{FW82,YAT96,XD07}, meaning $\sim$10$^{28}$ W is available to drive the wind: substantially more than the $\gtrsim$5.6 $\times$ 10$^{23}$ W mentioned at the start of Section \ref{DrivingSect}.

Of the energy expended by the stellar pulsation, almost all of it is required to overcome the gravitation potential of the star. The pulsation-enhanced, dust-driven wind model already relies on pulsation to provide half of this energy to levitate material from the stellar surface to $\sim$2 R$_\ast$, so it is not unreasonable to suppose it can provide the remaining half as well. Radiation pressure would then become a small additional component, creating a pulsation-P{\it driven}, dust-{\it enhanced} wind. 

Expansion velocity profiles of luminous evolved stars are not consistent with this pulsation-driven scenario (e.g.\ \citealt{REG+12}). However, these stars have greater mass-loss rates ($\sim$10$^{-5}$ M$_\odot$ yr$^{-1}$) and faster wind velocities ($\sim$10--30 km s$^{-1}$) than we observe here. They are typically either massive supergiants with weaker, stochastic pulsations, or heavily obscured AGB stars where trapped radiation can be more effective in driving a wind. While radiation pressure on winds may therefore be effective in these environments, they may not be effective in strongly pulsating but less obscured stars like EU Del. Given the difficulty of the other methods to drive a wind from EU Del, we suggest the dominant driving mechanism for the wind of this low-luminosity AGB star is the pulsation itself.



\section{Discussion}
\label{DiscSect}

\subsection{Comparison to VY Leo}

\begin{center}
\begin{table*}
\caption{Stellar parameters of EU Del compared to VY Leo.}
\label{StarCompTable}
\begin{tabular}{l@{}cc@{}r}
    \hline \hline
Parameter  & EU Del  & VY Leo & Ref. \\
    \hline
Distance ($d$)		& 117 $\pm$ 7 pc			& 119 $\pm$ 5 pc 		& 1\\
Tangential velocity	& 38 km s$^{-1}$			& 14 km s$^{-1}$		& 1\\
Radial velocity		& --66 km s$^{-1}$			& --8.39 km s$^{-1}$ 		& 2\\
Spectral type		& M6					& M5.5				& 2\\
Effective temperature
\rlap{($T_{\rm eff}$)}	& 3243 K				& 3300 K			& 3 \\
\			& 3227 K				& 3311 K			& 4\\
\			& \ 					& 3300 K			& 5\\
Luminosity ($L$)	& 1585 L$_\odot$			& 1357 L$_\odot$		& 4\\
\			& \ 					& 1346 L$_\odot$		& 5\\
log($g$)		& 0.39					& 1.00				& 6,3 \\
\ 			& \ 					& 0.20				& 7 \\
Radius ($R$)		& 127.5 R$_\odot$			& 112.1 R$_\odot$		& 4 \\
Escape velocity 
\rlap{($M=0.6$\,M$_\odot$)}	& 42 km s$^{-1}$			& 45 km s$^{-1}$		& \\
$W_{\rm JK}$ at LMC	& 11.20 mag				& 11.56 mag			& $^\dag$\\
Variability type	& SRV (O-rich)				& Irregular/SR			& 8,9 \\
Pulsation period	& 60.8, 132.6, 235.3, 67.3, 44.0 days	& 54.9, 84.5, 75.0, 35.8 days	& 10 \\
Amplitude (optical)	& 149, 113, 91, 75, 61 mmag		& 162, 76, 69, 69 mmag		& 10 \\
Pulsation mode		& C$^{\prime}$, C, ?, C$^{\prime}$, D/E?& C$^{\prime}$, F?, F?, B?	&  \\
Derived mass		& 1.80, {\it 0.76}, 0.40, 1.61, 2.59 M$_\odot$& 1.53, {\it 0.95}, 1.08, 2.46 M$_\odot$& 11$^\ast$ \\
Outflow velocity	& 9.505 $\pm$ 0.021 km s$^{-1}$		& 12.00 $\pm$ 0.35 km s$^{-1}$	& 11,12 \\
CO (2--1) flux		& 76.3 $\pm$ 1.5 Jy km s$^{-1}$		& 8.1 $\pm$ 0.5 Jy km s$^{-1}$	& 11,12 \\
CO (3--2) flux		& 146.5 $\pm$ 3.3 Jy km s$^{-1}$	& 4.9 $\pm$ 0.8 Jy km s$^{-1}$	& 11,12 \\
Mass-loss rate		& few $\times 10^{-8}$ M$_\odot$ yr$^{-1}$	& few $\times 10^{-9}$ M$_\odot$ yr$^{-1}$	& 11,12 \\
    \hline
\multicolumn{3}{p{0.93\textwidth}}{References: (1) \citet{vanLeeuwen07}; (2) \citet{AF12}; (3) \citet{CPI+13}; (4) \citet{MZB12}; (5) \citet{Groenewegen12}; (6) \citet{WSP+11}; (7) \citet{IRE+04}; (8) \citet{MMAL01}; (9) \citet{Watson06}; (10) \citet{TBK+09}; (11) this work; (12) \citet{Groenewegen14}. $^\dag W_{\rm JK}$ is a reddening-free Wesenheit index in the 2MASS $J$ and $K_{\rm s}$ bands. Here it is given \emph{at the distance of the Large Magellanic Cloud} (LMC; 50 kpc) for easy comparison to \citet{Wood15}. $^\ast$Derived masses are based on the period--mass--radius relationship of \citet{Wood90} for the different periods found by \citet{TBK+09}. Italic type indicates the adopted mass based on our assignment of the fundamental-mode period.}\\
    \hline
\end{tabular}
\end{table*}
\end{center}

\begin{figure}
\centerline{\includegraphics[height=0.47\textwidth,angle=-90]{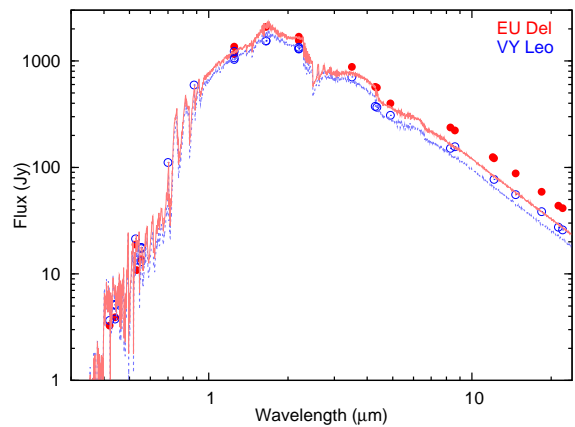}}
\caption{Spectral energy distribution of EU Del and VY Leo (large red and blue points, respectively). EU Del is shown with a {\sc bt-settl} model atmosphere at 3300 K, log($g$) = 0.5 dex, [Fe/H] = --0.5 dex (red line). The model for VY Leo is at 3200 K, log($g$) = 1.0 dex, [Fe/H] = 0 dex (dashed blue line). See text for the sources of the data.}
\label{SEDCompareFig}
\end{figure}

Few other giant stars of similar luminosity have been observed in these CO transitions. The closest literature comparison to EU Del is VY Leo (HIP 53449). Table \ref{StarCompTable} compares the properties of the two stars, and Figure \ref{SEDCompareFig} shows their SEDs. VY Leo is probably a little less evolved, having a slightly higher temperature and lower luminosity, giving it a slightly smaller radius. The photometry for these SEDs is derived from: \emph{Hipparcos} \citep{Perryman97,vanLeeuwen07}, the \emph{Cosmic Background Explorer / Diffuse Infrared Background Experiment} (COBE/DIRBE; \citealt{PSK+10}), the American Association of Variable Star Observers All-Sky Photometric Survey (APASS; \citealt{HLT+12}), \citet{JMIW66}, \citet{MM78}, \citet{Warren91}, the Two Micron All Sky Survey (2MASS; \citealt{SCS+06}), \emph{Akari} \citep{KAC+10}, the \emph{Wide-field Infrared Survey Explorer} (WISE; \citealt{CWC+12}), the \emph{Infrared Astronomical Satellite} (IRAS; \citealt{NHvD+84}), and the \emph{Midcourse Space Experiment} (MSX; \citealt{EP96}). As with EU Del, reddening to VY Leo has been neglected. Based on the extinction maps of \citet{LVV+14}, $E(B-V)$ is likely to be $<0.01$.

Spectroscopically, VY Leo is very similar to EU Del, although VY Leo shows a much more modest infrared excess \citep{Groenewegen12}. This is commensurate with the lower mass-loss rate derived from its fainter CO lines. VY Leo pulsates with a slightly shorter period. Notably, VY Leo does not have significant power in longer-period modes, while EU Del does. Power in shorter-period pulsation modes is typically associated with stars of a higher mass \citep[e.g.][]{Wood15}.

The log($g$) estimates disagree to the extent that it is impossible to define which is the more massive star. However, stellar mass can be estimated from the pulsation period via the period--mass--radius relationship of \citet{Wood90}:
\begin{equation}
\log P = -2.07 + 1.94 \log R - 0.9 \log M ,
\label{WoodEq}
\end{equation}
where period is in days, and radius and mass are in solar units. This calculation refers to the fundamental mode period. Table \ref{StarCompTable} lists the resulting masses if each pulsation mode was adopted as the fundamental mode. The italic type (0.76 M$_\odot$ for EU Del and 0.95 M$_\odot$ for VY Leo) indicates the mass corresponding to our identification of the fundamental mode for each star.

Table \ref{StarCompTable} also lists our identification of the pulsation modes for EU Del and VY Leo, based on the positions of these modes in the period--luminosity diagram of \citet{Wood15}. We use labels for the period--luminosity sequences from \citet{Wood15}, where sequence C is generally accepted as the fundamental mode, and C$^\prime$, B and A are overtone modes. The origins of sequence F are not clear, but \citet{Wood15} suggests this is the higher-mass counterpart of the fundamental mode. 

EU Del appears to be an overtone (sequence C$^\prime$) pulsator, with the fundamental period being 132.6 days. The fundamental and overtone modes are shifted to the shorter-period side of sequences B and C, as is normal for an overtone pulsator (see \citealt{Wood15}, his figure 4).

VY Leo is harder to place. The fundamental mode is not clear, as no mode in sequence C seems to be excited. Two modes bracketing sequence B are excited, as are two modes within the sequence F region. If sequence F represents the fundamental mode, a mass of $\sim$0.95 or 1.08 M$_\odot$ is implied for VY Leo. We have highlighted the mass corresponding to the slightly stronger 84.5-day period (0.95 M$_\odot$) in Table \ref{StarCompTable}. This would make VY Leo a more massive star.

For stars of the same luminosity, an observational correlation exists between power in longer-period, fundamental modes and the red infrared colours which indicate strong stellar mass loss \citep[][their figure 17]{BMS+15}. At the same luminosity, more massive stars have shorter-period pulsations. This is true of both their fundamental pulsation period (Eq.\ \ref{WoodEq}, \citealt{Wood90}) and mode in which they pulsate \citep[][their table 2]{FW82}. Therefore, at a given luminosity, longer pulsation periods imply lower stellar masses and higher mass-loss rates. Hence, strong mass loss will start earlier in lower-mass stars, driven by their long-period pulsations. Also, at any particular luminosity, a low-mass star will have stronger mass-loss than its high-mass equivalent.

In comparing EU Del and VY Leo, the increase in mass-loss rate appears to be caused by the more effective long-period pulsations driving mass loss. Hence this provides additional evidence that the wind of EU Del is driven by pulsations. Pulsation likely does not enhance the wind of VY Leo, hence it has a lower mass-loss rate. {While EU Del does not appear to be supported by a magnetically driven wind (Section \ref{MagWind}), VY Leo may be. Whatever the driving mechanisms involved,} the similarity between the outflow velocities of EU Del and VY Leo adds support to our claim that radiation pressure on dust is ineffective.



\subsection{Testable predictions for stellar mass loss}

We can now turn to other samples to examine the hypothesis of a pulsation-driven dust-enhanced wind. Pulsation amplitude and period are strong functions of effective temperature (\citealt{KB95} and Eq.\ \ref{WoodEq}, respectively). Effective temperature is a function of atmospheric opacity and hence metallicity. Thus, if pulsation amplitude and period are linked to mass-loss rate, we would expect pulsation-driven winds to depend strongly on metallicity. However, the mass-loss rates of metal-poor RGB stars at these luminosities are largely metallicity independent \citep{MZ15b}. This independence implies that pulsation-driven winds are not important on the RGB, and that the transition to a dust-producing, pulsation-driven wind only happens on the AGB. This would be observable in the luminosity distribution of dusty sources across the RGB tip. While the luminosity function of all stars decreases as one moves to brighter stars above the RGB tip, the luminosity function of AGB stars should not. This is tentatively observed in globular clusters already \citep[e.g.][]{MBvL+11,MSS+12}.

Stellar wind theory dictates that, in a pulsation-enhanced, dust-driven wind, the mass-loss rate is set by the pulsations, but the terminal velocity is set by the radiation pressure on dust (e.g., \citealt{LC99,NL02}). Numerous observations show that more luminous, dustier stars have faster winds, but that few single stars have CO lines with velocities below $\sim$5--10 km s$^{-1}$ (e.g.\ \citealt{Young95,KYLJ98}).

Few direct observations of terminal wind velocities exist in single stars without pulsation-enhanced winds, due to their low mass-loss rates and lack of condensed molecules. There are indirect measurements from asymmetries in chromospherically active lines \citep{MvL07,MDS08,MDS09}. The absorption cores of these lines form above the stellar chromosphere, but typically before the winds have reached their terminal velocity. They typically show a velocity of a few km s$^{-1}$ just above the stellar surface, again increasing with luminosity. Near-IR He {\sc i} line observations of early AGB stars indicate a much faster terminal velocity ($\sim$30--100 km s$^{-1}$; \citealt{DSS09}), implying that the terminal velocity should \emph{decrease} with increasing luminosity in the early-AGB phase, but these are difficult measurements that are hard to confirm.

We can therefore expect that terminal wind velocities of early-AGB stars are either stable and low, or decline with luminosity. There should then exist an inflection point, or a velocity minimum, after which the wind velocities increase in response to dust driving. This point should be reached at higher luminosities in metal-poor stars. EU Del and VY Leo should straddle this inflection, which probably occurs near 10 km s$^{-1}$. Further measurements of early-M-type stars likely to occupy this transition region would determine when dust driving is likely to begin. Identification of this point will require accurate luminosities for these sources, which will require accurate distances.


\section{Conclusions}
\label{ConcSect}

We have observed the CO outflow from EU Del, finding a wind close to the canonical 10 km s$^{-1}$ but with a mass-loss rate of a few $\times$ 10$^{-8}$ M$_\odot$ yr$^{-1}$. The metallicity and kinematic origin of EU Del require further refinement. We project it to a relatively metal-poor member of the thin disk population of our Galaxy, although the uncertainty on the metallicity is significant. Using physical arguments, and by comparison to the similar star VY Leo, we suggest that the wind of EU Del is largely driven by stellar pulsation. Radiation pressure on dust does not appear to modify the outflow velocity. We suggest that EU Del is a low-mass AGB star that has recently started strong enough mass loss to initiate dust production, thanks to excitation of its 132.6-day, fundamental-mode pulsation. We encourage further observations of CO and dust production around stars near the RGB tip to identify the link between pulsation, stellar mass and mass-loss rate.


\section*{Acknowledgements}

This publication is based on data acquired with the Atacama Pathfinder Experiment (APEX). APEX is a collaboration between the Max-Planck-Institut fur Radioastronomie, the European Southern Observatory, and the Onsala Space Observatory. Based on observations made with ESO telescopes under programme IDs 092.F-9328 and 094.F-9328. C.I.J.\ gratefully acknowledges support from the Clay Fellowship, administered by the Smithsonian Astrophysical Observatory. The authors are extremely grateful to Sofia Ramstedt, who reduced the data for this project, both for her work on the data and for her insightful comments during the preparation of this manuscript.

\label{lastpage}

\end{document}